\documentclass[twocolumn,showpacs,preprintnumbers,amsmath,amssymb]{revtex4}
\usepackage{epsfig}
\usepackage{amsmath}

\begin{document}

\title{Detecting Neuronal Communities from Beginning of Activation Patterns}

\author{Luciano da Fontoura Costa}
\affiliation{Institute of Physics at S\~ao Carlos, University of
S\~ao Paulo, PO Box 369, S\~ao Carlos, S\~ao Paulo, 13560-970 Brazil}

\date{27th Jan 2008}

\begin{abstract}
The detection of neuronal communities is addressed with basis on two
important concepts from neuroscience: facilitation of neuronal firing
and nearly simultaneous beginning of activation of sets of neurons.
More specifically, integrate-and-fire complex neuronal networks are
activated at each of their nodes, and the dissemination of activation
is monitored.  As the activation received by each neuron accumulates,
its firing gets facilitated.  The time it takes for each neuron, other
than the source, to receive the first non-zero input (beginning
activation time) and the time for it to produce the first spike
(beginning spiking time) are identified through simulations.  It is
shown, with respect to two synthetic and a real-world
(\emph{C. elegans}) neuronal complex networks, that the patterns of
beginning activation times (and to a lesser extent also of the spiking
times) tend to cluster into groups corresponding to communities of
neurons in the original complex neuronal network.  Such an effect is
identified to be a direct consequence of the almost simultaneous
activation between the nodes inside the same community in which the
source of activation is placed, as well as of the respective trapping
of activation implied by the integration of activiation prior to
firing.  Interestingly, the accumulation of activity and thresholds
inside each neuron were found to be essential for constraining the
initial activations within each respective community during the
transient activation (no clear clusters were observed when using
overall activation or spiking rates).  In addition to its intrinsic
value for neuroscience and structure-dynamics studies, these results
confirm the importance of the consideration of transient dynamics in
complex systems investigations.
\end{abstract}

\maketitle

\vspace{0.5cm}
\emph{`The brain is a wonderful organ. It starts working the moment you 
get up in the morning and does not stop until you get into the office. 
(R. Frost)}

\section{Introduction} 

Much has been investigated about neuronal systems from both the
biological and exact sciences points of view
(e.g.~\cite{Squire:2003}).  More recently, neuronal networks met
complex networks (e.g.~\cite{Stauffer_Hopfield, Stauffer_Costa,
Costa_revneur:2005, Kim:2004, Timme:2006, Osipov:2007, Hasegawa:2004,
Hasegawa:2005, Park:2006}).  Such an interface is particularly
promising, as it allows the emphasis of dynamical systems
characterizing research in neuronal networks to be integrated with the
structural approaches of complex networks research
(e.g.~\cite{Albert_Barab:2002, Newman:2003, Dorogov_Mendes:2002,
Costa_surv:2007}).  The intersection of these two major areas, which
lies at the heart of the structure and dynamics relationship
(e.g.~\cite{Newman:2003, Boccaletti:2006}), is henceforth called
\emph{Complex Neuronal Networks research}.  However, despite the growing 
attention to this are, few works have considered simple neuronal
models such as the integrate-and-fire.  In addition, rather few
studies have addressed transient dynamics
(e.g.~\cite{Costa_Ising:2007}) or the accumulation of stimuli
responsible for the facilitation of firing~\cite{Squire:2003}.

In a recent study~\cite{Costa_nrn:2008}, the transient dynamics of
integrate-and-fire networks underlain by several types of connectivity
was characterized with respect to a series of dynamical properties,
including the activation of nodes, spiking, and onset times for
activation and spiking.  As the activation received by each neuron was
stored inside it as its state (facilitation~\cite{Squire:2003}), it
became possible to investigate the activation and spiking separately.
The neuronal dynamics was found to vary markedly with respect to the
connectivity, with abrupt transitions of initiation of generalized
spiking being observed for some complex networks models, as well as
for the \emph{C. elegans} network. The current work continues such a
investigation in order to investigate for possible simultaneous
neuronal activation as a consequence of concentrated interconnectivity
between groups of neurons.  The basic idea is that more intensely
interconnected groups of neurons, i.e. communities of nodes
(e.g.~\cite{Watts_Strogatz:1998, Girvan:2002, Zhou:2003, Newman:2004,
Radicchi:2004, Hopcroft:2004, Latapy:2005, Guimera:2005, Bagrow:2005,
Capocci:2005, Arenas:2008, Costa_comm:2006}), would imply
concentration of the accumulated activation because of the thresholds,
delaying the dissemination of the activation to other parts of the
network.  More specifically, external activation is fed into each of
the $N$ neurons, and the unfolding activation is monitored.  The
beginning activation time of each neuron $i$ (except that used as
source) is experimentally determined, corresponding to the time from
the beginning of the external activation to the first arrival of
non-zero stimuli at neuron $i$.  Therefore, a pattern of $N-1$
beginning activation times are obtained for each place of stimulation.
Because of the high dimensionality of these patterns, the statistical
method known as Principal Component Analysis --- PCA
(e.g.~\cite{Costa_book:2001, Costa_surv:2007}) is applied for
dimensionality reduction (through optimal decorrelation), allowing the
visualization of the distribution of patterns of beginning activation
times.  Interestingly, these patterns were found to be organized in
well-defined clusters for two synthetic networks, allowing the
immediate identification of the neuronal communities.  Structure
distributions, also containing clusters, was also observed for the
\emph{C. elegans} network. Similar, though a little less effective,
results were obtained by considering the beginning spiking times.  No
clusterings are observed in case the activation of spiking patterns at
specific times (or even averages) are taken into account.  Such
results corroborates the importance of transient dynamics in complex
systems investigations.

This article starts by presenting the basic concepts in complex
networks, integrate-and-fire complex neuronal networks and the
statistical method of Principal Component Analysis.  The results,
discussion and perspectives for future investigations are presented
subsequently.

\section{Basic Concepts}

A directed, unweighted network $\Gamma$ can be fully represented in
terms of its \emph{adjacency matrix} $K$.  Each edge extending from
node $i$ to node $j$ implies $K(j,i) = 1$.  The absence of connection
between nodes $i$ and $j$ is represented as $K(j,i)=0$.  The
\emph{out-degree} of a node $i$, henceforth expressed as $k$,
corresponds to the number of outgoing edges of that node.  An analogue
definition holds for the in-degree. The \emph{immediate neighbors} of
a node $i$ are those nodes which can be reached from $i$ through an
outgoing edge.  Two nodes are \emph{adjacent} if they share an edge;
two edges are adjacent if they share a node.  A \emph{walk} is a
sequence of adjacent edges, with possible repetitions of nodes and
edges.  A \emph{path} is a walk without repetition of nodes or edges.
The \emph{length} of a walk (or path) is equal to the number of the
edges it contains.  The \emph{shortest path length} between two nodes
is the length of the shortest path between them.

An \emph{integrate-and-fire neuron} involves two successive stages:
(i) summation of the inputs $x_i$, i.e. $S = \sum_{i} x_i$; and (ii)
thresholding, i.e. a spike is produced whenever $S$ is larger than a
threshold $T$.  In this work, each node represents an
integrate-and-fire neuron.  The incoming activation (from dendrites)
is accumulated as its state until a spike occurs, in which case the
accumulated activation is completely flushed out through the outgoing
connections (axons).  In order to ensure conservation of the total
activation, ach outgoing axon conveys a fraction $S/k(i)$ of the
respective previously accumulated activation, where $k(i)$ is the
out-degree of neuron $i$.  The activation and spiking of all neurons
in the network can be represented through the respective
\emph{activogram} and \emph{spikegram}, namely matrices storing the 
activation or occurrence of spikes for every node along all considered
times.  In this article, all neurons have the same threshold $T=1$,
and the external activation of the network is always performed by
injecting activation of intensity $1$ at each of the neurons.  For
each of these activations (with the source of activation placed at
node $v$), the time one neuron $i$ takes, from the beginning of the
external initiation, to receive the first non-zero input is henceforth
called its respective ~\emph{beginning activation time} $T_a(i,v)$.
The time it takes for that neuron to produce the first spike is
the~\emph{beginning spiking time} $T_s(i,v)$.

The dimensionality of the measurement space defined by the onset
activation times for each node in a network involves $N$
measurements~\footnote{Though a total of $N-1$ non-zero beginning
times is obtained for each activation, we consider a vector of $N$
measurements by incorporating the zero time respective to the
activation source.} (the beginning times).  So, we have a total of $N$
measurements for each of the $N$ nodes, which is a high dimensional
space involving several correlations between the times.  The
dimensionality of such measurements spaces can be convenient and
optimally reduced (decorrelation) by using the Principal Component
Analysis (PCA) methodology (e.g.~\cite{Costa_book:2001,
Costa_surv:2007}).  Let each of the $N$ observations $v =
\{1, 2,\ldots, N \}$, characterized by the set of beginning times at 
each neuron $i$ as a consequence of activation placed at node $v$, be
organized as respective \emph{feature vectors} $\vec{f_v}$, with
respective elements $f_v(i) = T_a(i,v)$, $i \in \{1, 2, \ldots, N\}$.
Let the \emph{covariance matrix} between each pair of measurements $i$
and $j$ be defined in terms of its elements

\begin{equation}
  C(i,j) = \frac{1}{N-1} \sum_{v=1}^{N} (f_v(i) - \mu_i)(f_v(j) - \mu_j)
\end{equation}

where $\mu_i$ is the average of $f_v(i)$ over the $N$ observations.
The eigenvalues of $C$, sorted in decreasing order, are represented as
$\lambda_i$, $i = 1, 2, \ldots, M$, with respective eigenvectors
$\vec{v_i}$.  The following matrix, obtained from the eigenvectors of
the covariance matrix, defines the stochastic linear transformation
known as the \emph{Karhunen-Lo\`eve
Transform}~\cite{Costa_book:2001,Costa_surv:2007}.

\begin{equation} \label{eq:PCA}
  G  =  \left [ \begin{array}{ccc}
              \longleftarrow  &  \vec{v_1}  &  \longrightarrow  \\
              \longleftarrow  &  \vec{v_2}  &  \longrightarrow  \\
              \ldots          &  \ldots     &  \ldots  \\
              \longleftarrow  &  \vec{v_m}  &  \longrightarrow  \\
             \end{array}   \right]  
\end{equation}

where $m=N$.  Because such a transformation concentrates the variance
of the observations along the first axes (the so-called principal
axes), it is frequently possible to reduce the dimensionality of the
measurements without losing much information (the variances along the
other axes tend to be small as a consequence of correlations between
the original measurements) by considering in the above matrix only the
$m < N$ eigenvectors associated to the larges eigenvalues. The new
measurements $\vec{g}$, with dimension $m$, can now be
straightforwardly obtained as

\begin{equation}
  \vec{g} = G \vec{f}.
\end{equation}

\section{Results and Discussion}

\begin{figure}[htb]
  \vspace{0.3cm} 
  \begin{center}
  \includegraphics[width=1\linewidth]{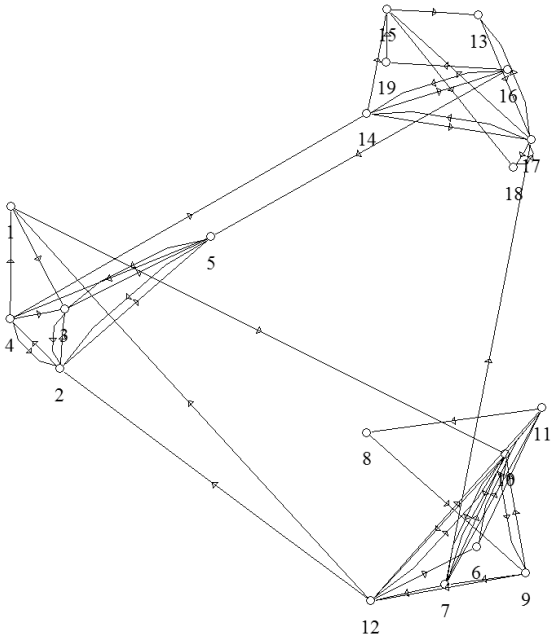} \\
   \caption{A simple network (\emph{Net1}) involving 3 communities
               with 5, 7 and 7 nodes considered in this work for 
               illustrative purposes.
  }~\label{fig:net1} \end{center}
\end{figure}

The potential of the neuronal community detection approach reported in
this article is illustrated with respect to the three following
directed networks: (a) a synthetic network (\emph{Net1}) containing 3
small communities (Figure~\ref{fig:net1}); (b) a medium-sized
synthetic network (\emph{Net2}) containing 4 communities
(Figure~\ref{fig:net1}); and (c) the network of
\emph{C. elegans} (\emph{NetCe})~\cite{Watts_Strogatz:1998}. 
The two synthetic communities were obtained by randomly assigning
direct edges among each of the communities, extracting the connected
component, and interconnecting the communities according to a fixed
probability.  Each of the three communities in \emph{Net1} contains 5,
7 and 7 nodes, respectively.  Each of the four communities in
\emph{Net2} contains 20, 37, 22 and 24 nodes.  The largest strongly
connected component in the \emph{C. elegans} network, used in this
work, contained 239 nodes.

\begin{figure*}[htb]
  \vspace{0.3cm} 
  \begin{center}
  \includegraphics[width=0.8\linewidth]{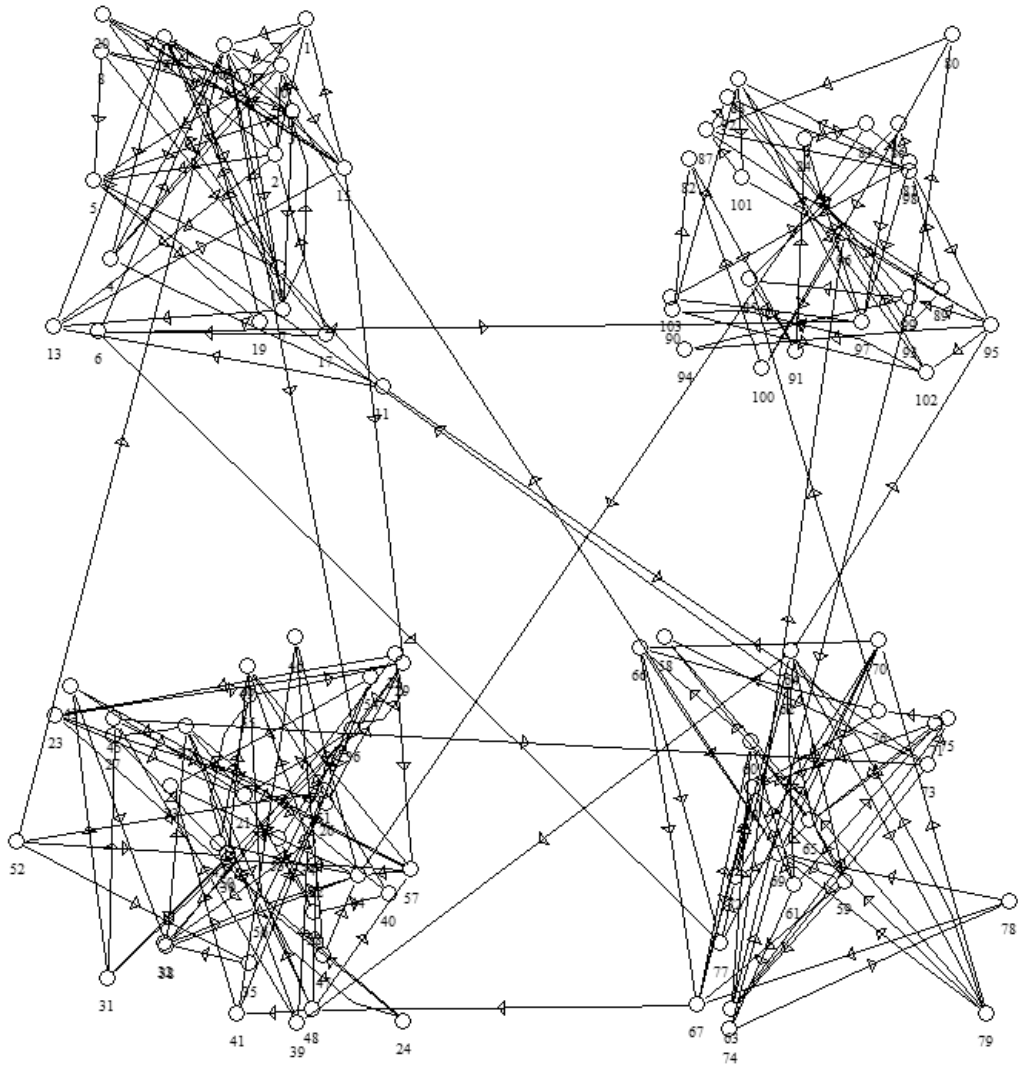} \\
   \caption{A medium-sized complex network (\emph{Net2}) containing 4 
                communities with 20, 37, 22 and 24 nodes each.
  }~\label{fig:net2} \end{center}
\end{figure*}

\begin{figure*}[htb]
  \vspace{0.3cm} 
  \begin{center}
  \includegraphics[width=0.3\linewidth]{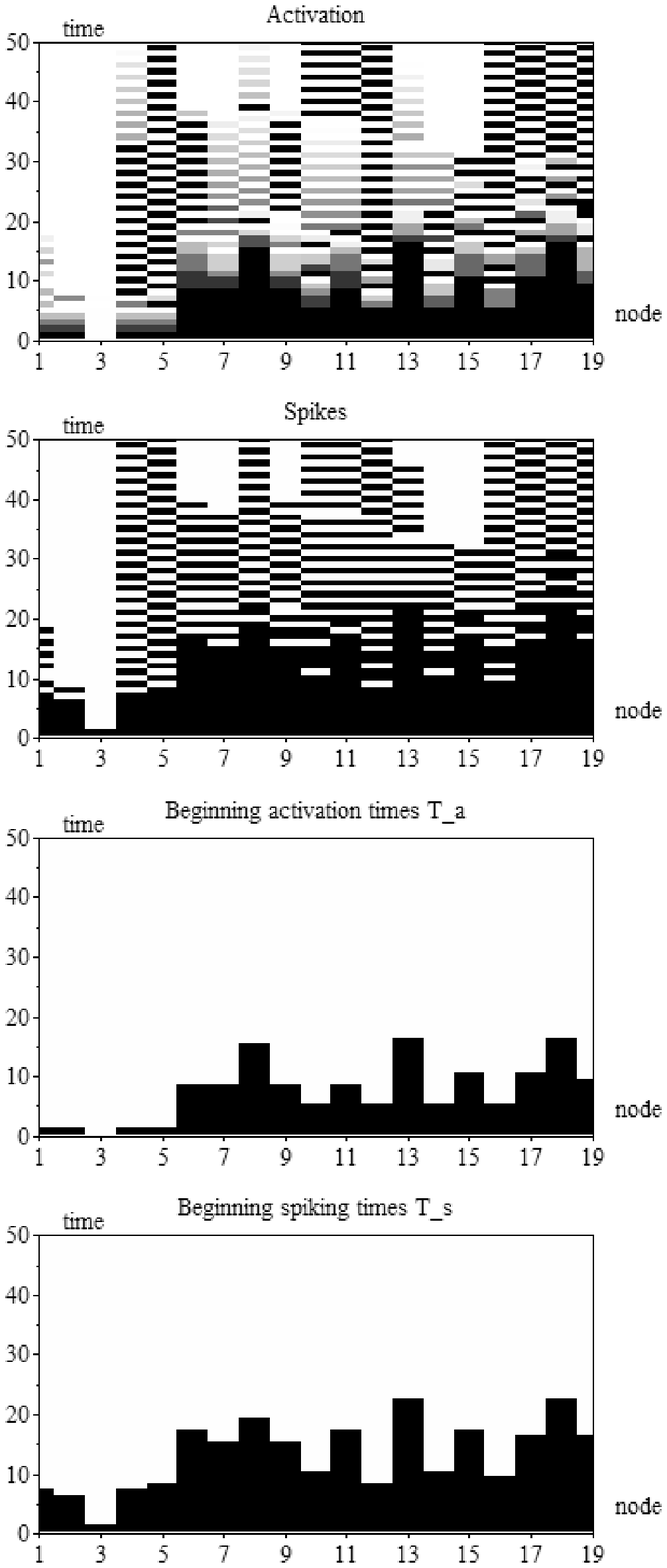} 
  \includegraphics[width=0.3\linewidth]{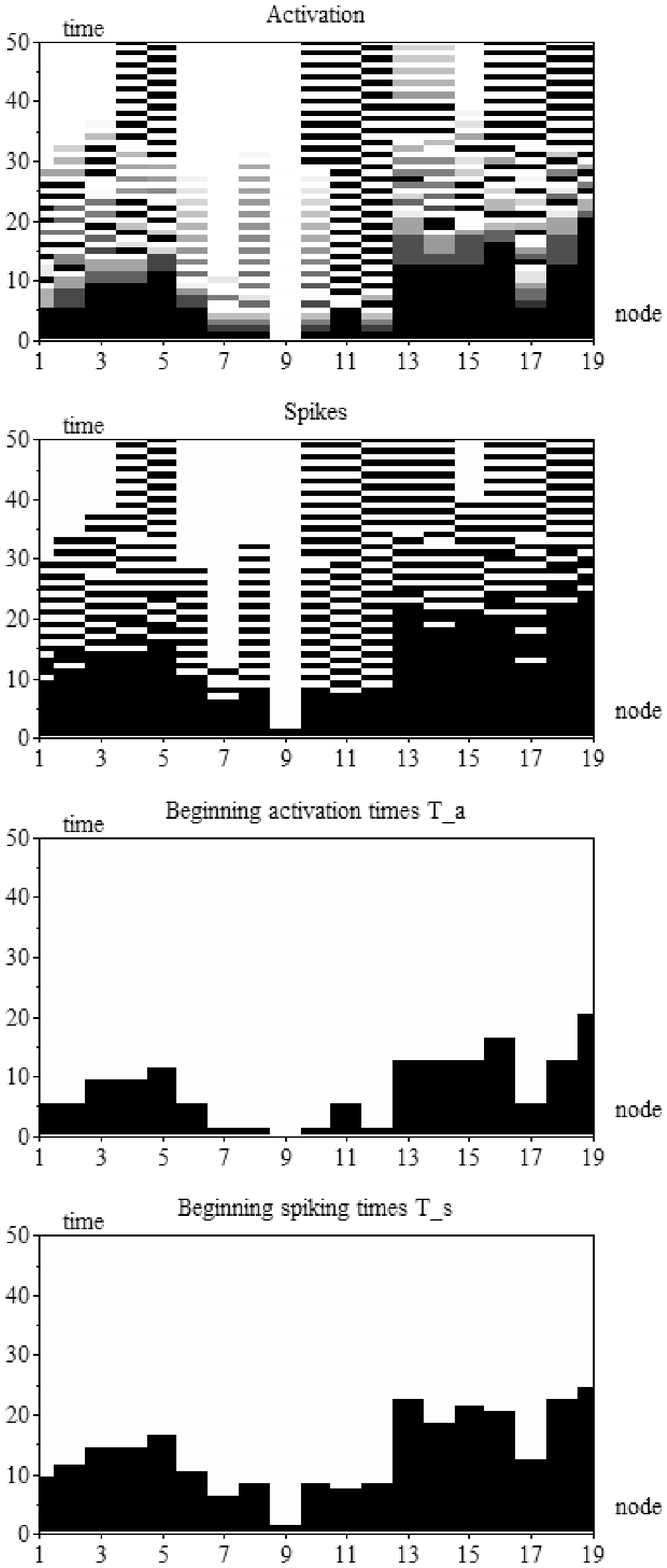} 
  \includegraphics[width=0.3\linewidth]{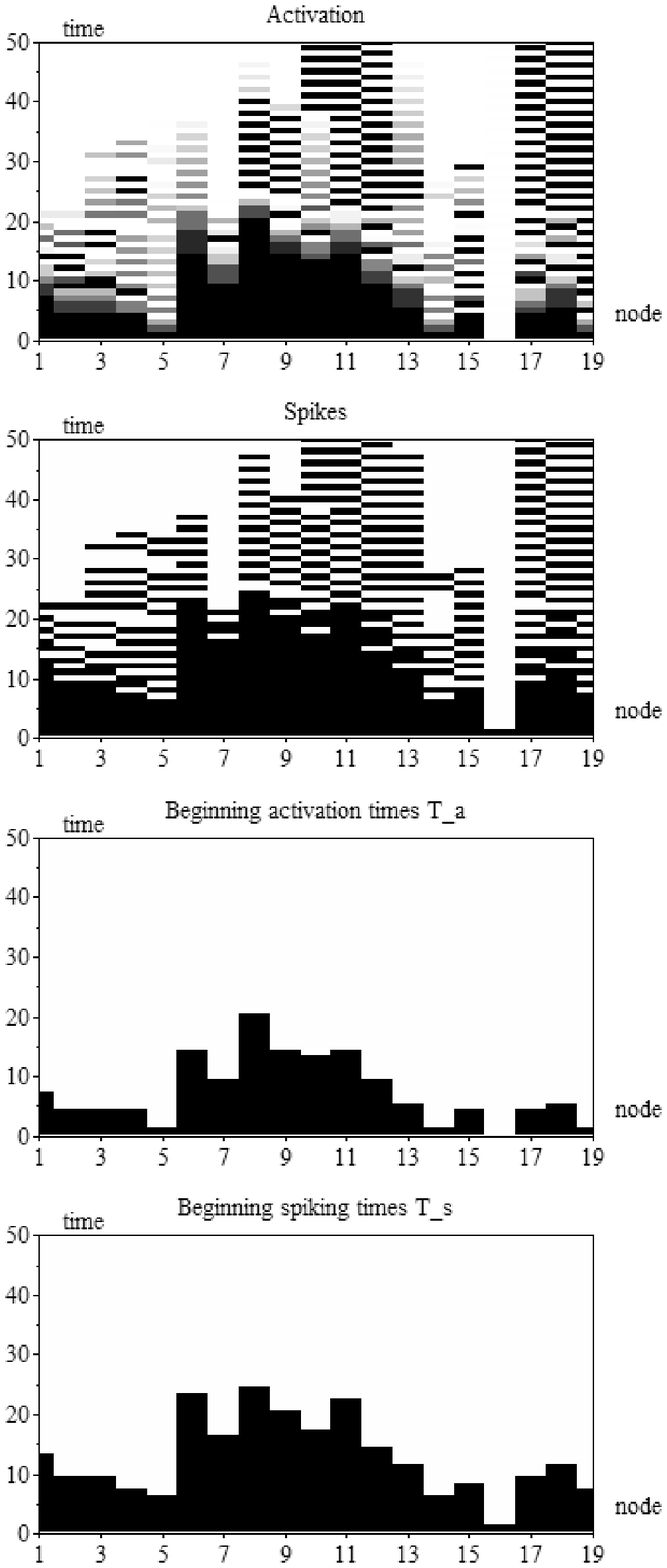}  \\
  (a) \hspace{5cm} (b) \hspace{5cm}  (c)
   \caption{The activograms and spikegrams, as well as the
               respective beginning activation and beginning
               spiking times, are shown with respect to the
               situations where the activation source corresponded
               to nodes 3 (a), 9 (b) and 16 (c).  The beginning
               time diagrams show in black the time instants 
               preceding the first activation or spiking of each
               neuron.  For instance, in the beginning spiking
               times diagram in (a), neuron 5 started
               spiking at the 8th time step from the initiation
               of the external activation arriving at node 3.
  }~\label{fig:grams} \end{center}
\end{figure*}

Figure~\ref{fig:grams} shows the activograms and the spikegrams, as
well as the respective beginning activation and beginning spiking
times diagrams for activation placed at node 3 (a), 9 (b) an 16 (c).
It is clear from the respective beginning activation time diagram in
Figure~\ref{fig:grams}(a) that the activation being received by node 3
implied in early and simultaneous conveyance of non-zero activation to
the other nodes in the community to which node 3 belongs (i.e. the
community including nodes 1 to 5).  Observe that the beginning spiking
times, shown in the respective diagram, are larger than the activation
time, because the respective neuronal firing requires accumulation of
the activation received by the dendrites of the neurons in that
community.  Similar nearly simultaneous activations of the other
communities can be identified in Figure~\ref{fig:grams}(b) and (c).
However, a less uniform initiation of activation in the this
community is observed in Figure~\ref{fig:grams}(c).

\begin{figure*}[htb]
  \vspace{0.3cm} 
  \begin{center}
  \includegraphics[width=0.8\linewidth]{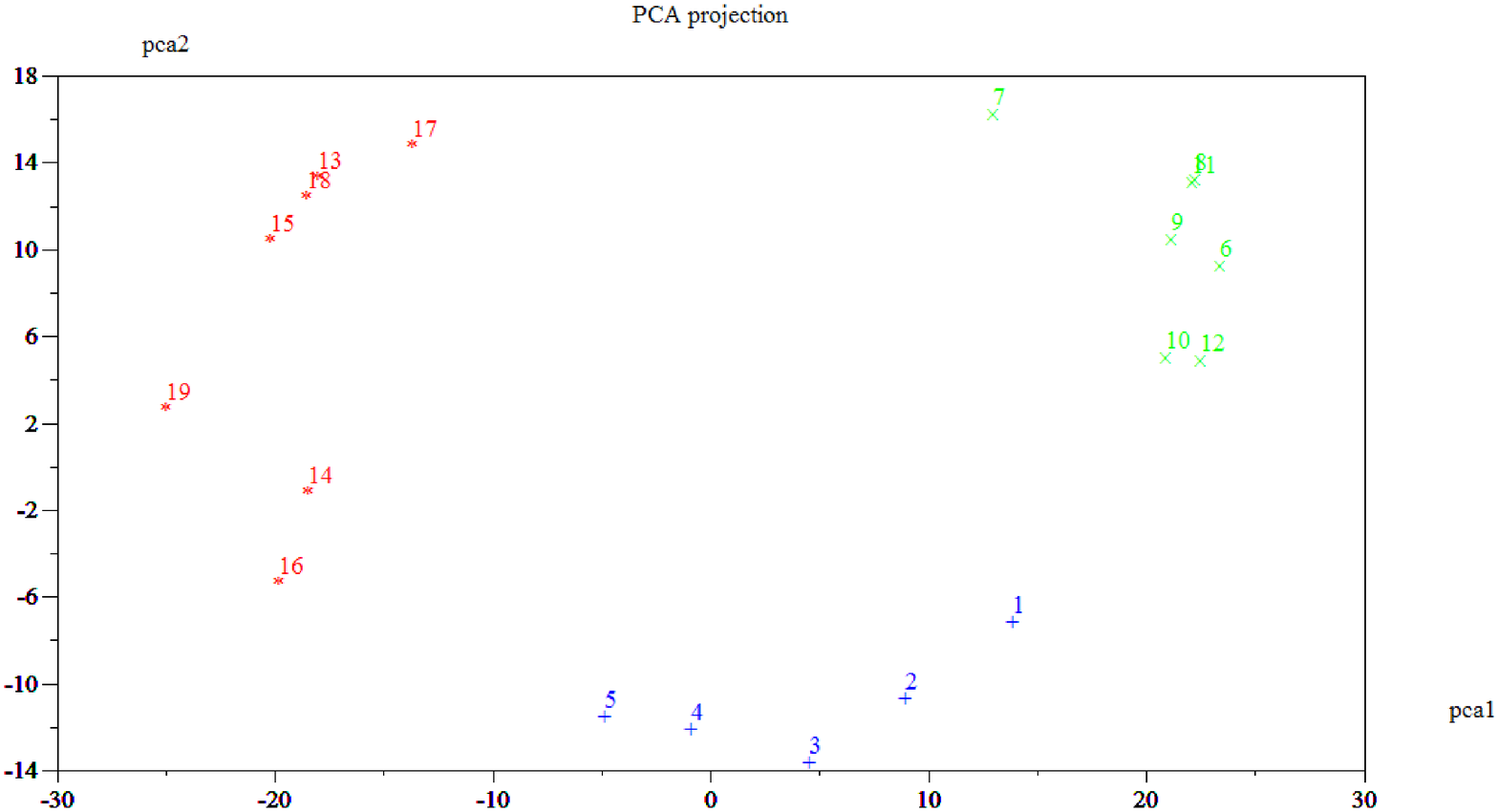} \\
   \caption{The PCA projection of the patterns of beginning activation 
                times obtained for \emph{Net1}.  Each of the three
                original communities can be clearly identified from the
                three respective clusters in this scatterplot.
  }~\label{fig:pca_net1} \end{center}
\end{figure*}

Figure~\ref{fig:pca_net1} shows the distribution of the beginning
activation times after PCA projection onto a two-dimensional space
defined by the principal variables $pca1$ and $pca2$.  The three
communities yielded well-defined respective clusters, which can be
immediately identified by using traditional clustering methods
(e.g.~\cite{Costa_book:2001}).  Observe that the nodes appearing at
the borders of each of the three clusters in Figure~\ref{fig:pca_net1}
correspond precisely to those nodes implementing the intercommunity
connections (see also~\cite{Guimera:2005}).  

\begin{figure*}[htb]
  \vspace{0.3cm} 
  \begin{center}
  \includegraphics[width=0.8\linewidth]{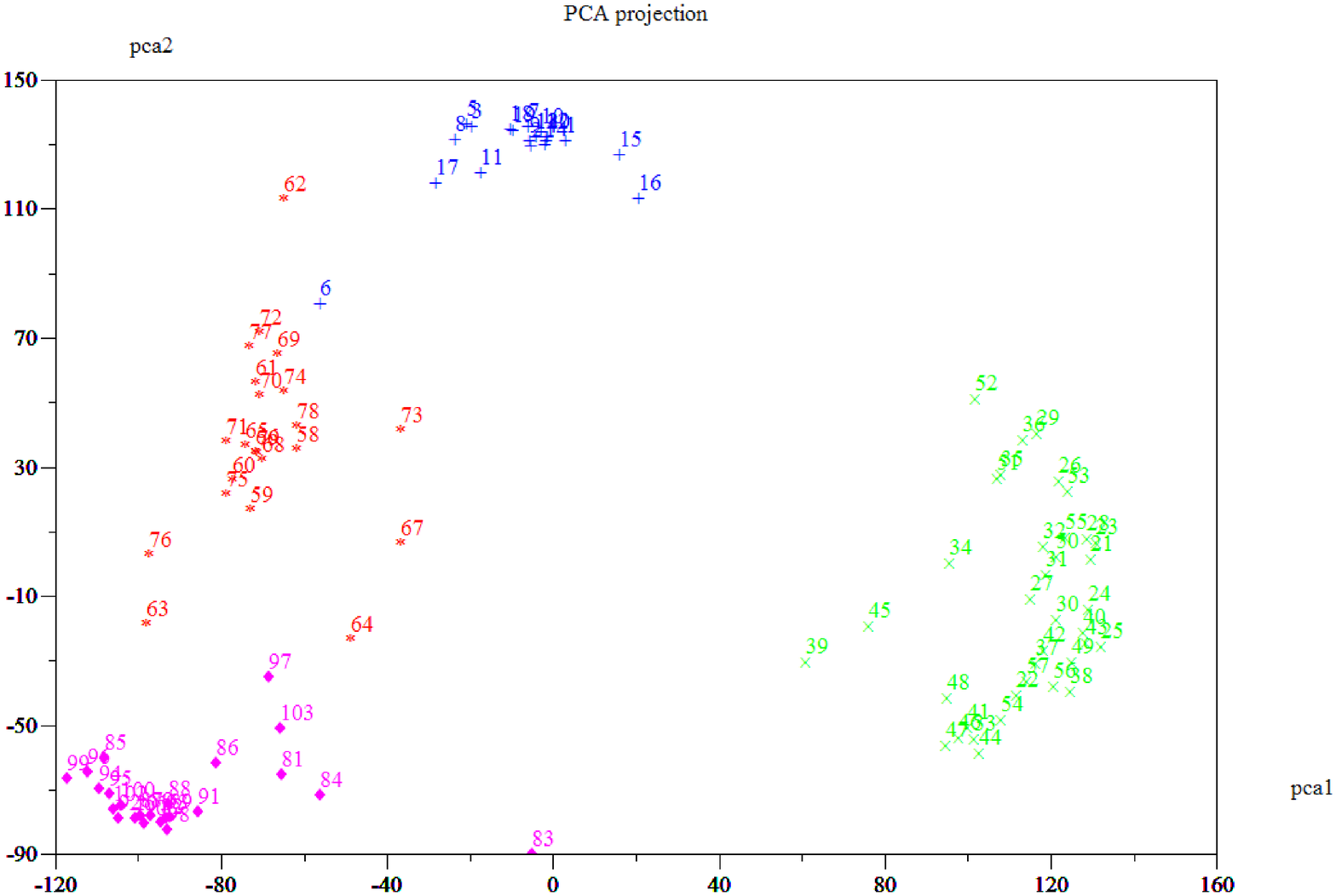} \\
   \caption{The PCA projection of the patterns of beginning activation 
                times obtained for \emph{Net2}.  Each of the four
                original communities can be clearly identified from the
                respective clusters in this scatterplot, with the
                nodes at the borders of the clusters corresponding to the
                nodes at the borders of the original communities.
  }~\label{fig:pca_net2} \end{center}
\end{figure*}

Figure~\ref{fig:pca_net2} depicts the clustering structure obtained
for network $Net2$.  Again, each of the communities was clearly mapped
into respective clusters in the two-dimensional PCA projected space.
Again, the bordering nodes in each cluster can be found to correspond
to the interface nodes between communities in the original network.
It is interesting to observe that, compared to the previous example,
the larger number of communities and nodes in this network tended to
imply a more cluttered distribution, especially at the interface
between the red and magenta communities.  Substantially more separated
clusters have been observed in three-dimensional PCA projected spaces.

\begin{figure*}[htb]
  \vspace{0.3cm} 
  \begin{center}
  \includegraphics[width=0.8\linewidth]{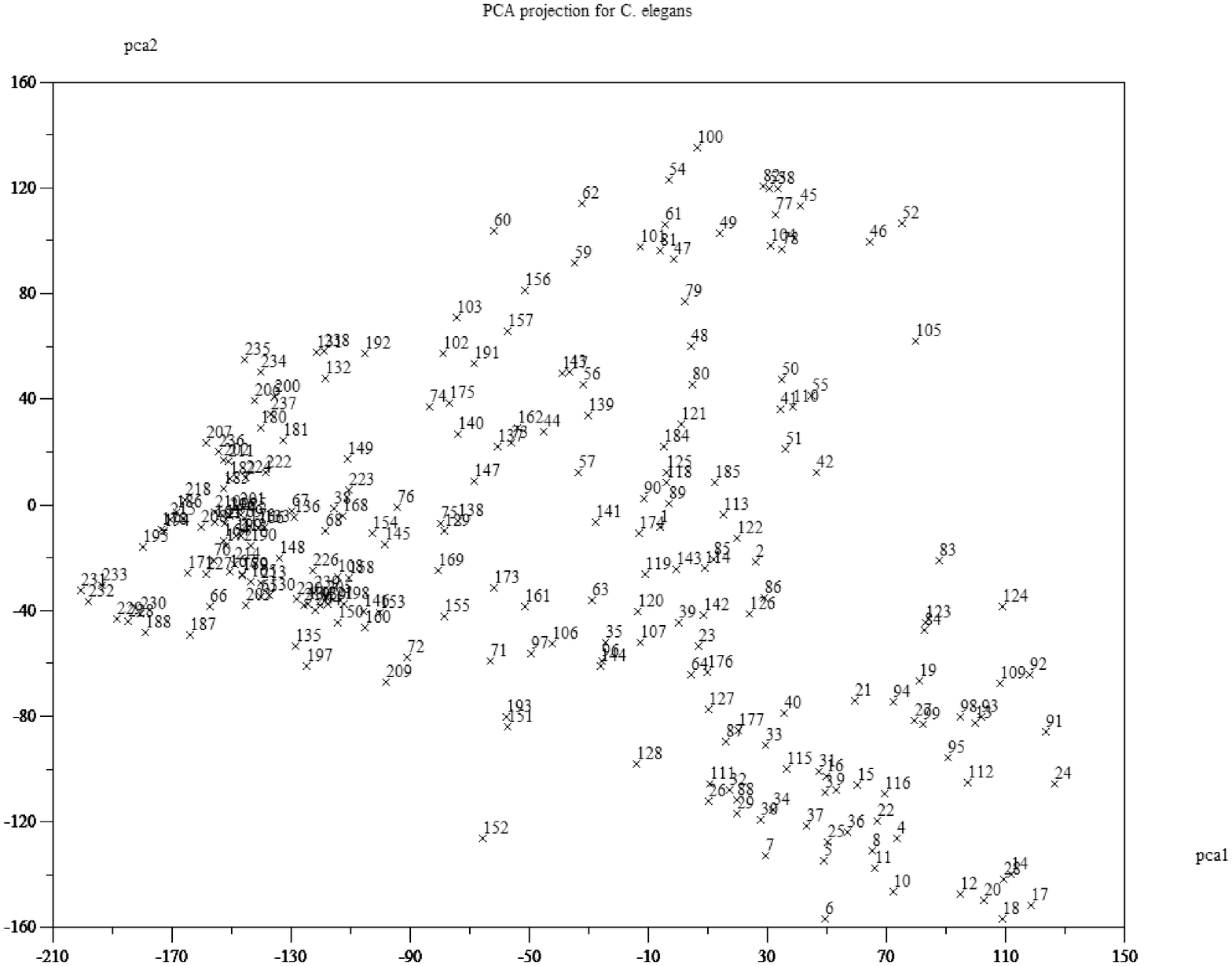} \\ 
  \caption{The distribution of nodes obtained for the \emph{C. elegans}
              network considering the two principal PCA variables
              $pca1$ and $pca2$.
  }~\label{fig:pca_Celegans} \end{center}
\end{figure*}

\begin{figure*}[htb]
  \vspace{0.3cm} 
  \begin{center}
  \includegraphics[width=0.8\linewidth]{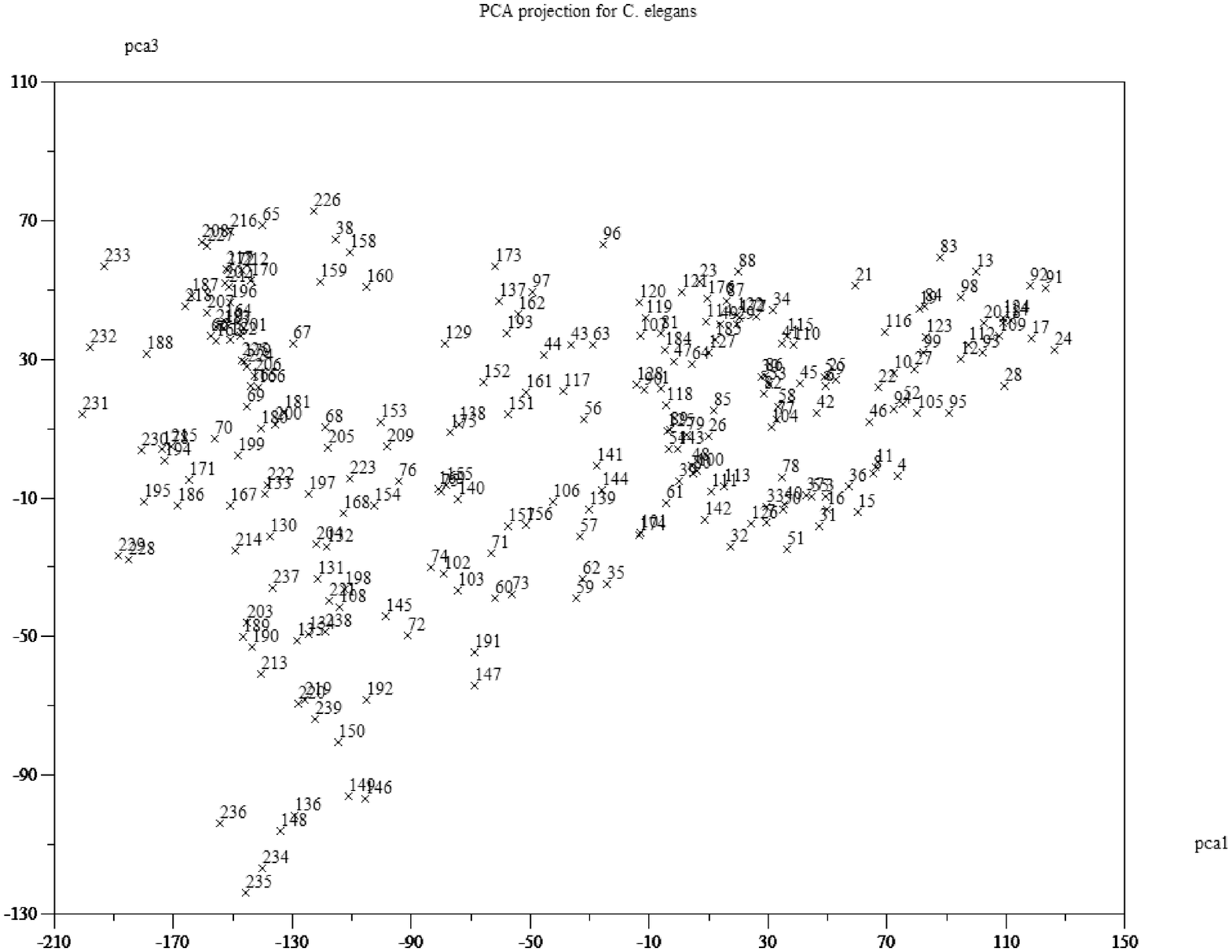} \\
   \caption{The distribution of nodes obtained for the \emph{C. elegans}
              network considering the first and third principal 
              PCA variables $pca1$ and $pca3$.
  }~\label{fig:pca_Celegans_z} \end{center}
\end{figure*}

The distribution of nodes obtained for the \emph{C. elegans} network
by two-dimensional PCA projection is shown in
Figure~\ref{fig:pca_Celegans}.  A concentration of nodes can be
observed at the left-hand side of the space, containing the nodes with
higher numbers (the numbers follow the original assignement as
in~\cite{Watts_Strogatz:1998}).  A community can also be discerned at
the lower right-hand side of the transformed measurement space.
Because of the large number of nodes in this network, it is
interesting to consider additional dimensions in the PCA projection.
The measurement space defined by the principal variables $pca1$ and
$pca2$ is shown in Figure~\ref{fig:pca_Celegans_z}.  This additional
projection shows that the denser cluster at the left-hand side of
Figure~\ref{fig:pca_Celegans} is actually scattered along the third
variable $pca3$, with a more compact cluster of nodes appearing at the
upper left-hand side of Figure~\ref{fig:pca_Celegans_z}.  In addition,
a small cluster involving nodes 136, 146, 148, 149, 234, 235 and 236
is now identifiable at the lower left-hand side of
Figure~\ref{fig:pca_Celegans_z}.

Less definite results were obtained in all cases by considering the
beginning spiking times, and no clear cluster structure was observed
when other activation of spiking measurements (at given instants or
averaged) were used.

\section{Concluding Remarks}

This work has addressed an important issue related to the
structure-dynamics paradigm in neuronal and complex networks research.
More specifically, we have investigated how communities of neurons in
directed complex neuronal networks can be identified by considering
the transient dynamics of beginning activation of nodes.  As a
consequence of the integration period required for reaching the firing
threshold in each integrate-and-fire neuron, the activation incoming
from the source node tends to be trapped inside the respective
community, unfolding to other portions of the network only after most
of the neurons in that community have started spiking.  The
distribution of the activation flushed outside each neuron at the
spikings, required for the conservation of the activation, was also
critical for the compartmentalization of the activation inside
communities.  The distinct patterns of beginning activation times
obtained by placing the activation source at each of the neurons of
each community were clearly revealed by the optimal statistical method
of Principal Component Analysis.  More specifically, the nodes tended
to cluster into respective groups, with the nodes at the borders of
such groups corresponding to those nodes implementing the
intercommunity connection in the original network.  In addition to its
intrinsic value for biological neuronscience, these results also
provide effective and simple practical means for obtaining neuronal
communities.

Several interesting future works are possible.  First, it would be
important to perform a more systematic and comprehensive study of the
separability of the communities by considering other types of
networks, with distinct interconnectivity between communities, among
other possibilities.  Also interesting is to use hierarchical
clustering methods (e.g.~\cite{Costa_book:2001, Zhou:2003,
Costa_surv:2007}) in order to obtain a hierarchical organization of
the neuronal communities, as well as investigating how such
hierarchies (e.g.~\cite{Arenas:2008}) are organized with respect to
time.  As the suggested method can be immediately extended to
identification in other types of networks, including non-directed
structures, it would be interesting to compare this method with other
more traditional approaches not involving thresholds.  Because abrupt
beginning of spiking has been observed~\cite{Costa_nrn:2008} for
several types of complex networks, it would be also interesting to
search for possible phase transitions of activation inside each
community, which could be ultimately responsible for the activation
trapping inside each neuronal community during the transient
activation period.

All in all, the findings and perspectives reported in this article
have supported the fact that investigations of transient non-linear
dynamics are specially promising and useful in the study of complex
systems (see also~\cite{Latora_entropy, Gardenes:2007,
Costa_diverse:2008, Costa_sync:2008, Costa_nrn:2008}).

\begin{acknowledgments}
Luciano da F. Costa thanks CNPq (308231/03-1) and FAPESP (05/00587-5)
for sponsorship.
\end{acknowledgments}

\bibliography{neucomm}
\end{document}